\documentclass[lettersize,journal]{IEEEtran}
\usepackage{amsmath,amsfonts}
\usepackage{algorithmic}
\usepackage{algorithm}
\usepackage{array}
\usepackage[caption=false,font=normalsize,labelfont=sf,textfont=sf]{subfig}
\usepackage{textcomp}
\usepackage{stfloats}
\usepackage{url}
\usepackage{verbatim}
\usepackage{graphicx}
\usepackage{cite}
\usepackage{multirow}
\begin{document}

\title{Volt-PF Control Mode for Distribution Feeder Voltage Management Under High Penetration of Distributed Energy Resources\\}

\author{Madhura~Sondharangalla,~\IEEEmembership{Student Member,~IEEE,}
        Dan~Moldovan,~\IEEEmembership{Student Member,~IEEE,}\\
        and~Raja~Ayyanar,~\IEEEmembership{Fellow,~IEEE}
\thanks{
\par Authors are with the Department of Electrical, Computer and Energy
Engineering, Arizona State University, Tempe, AZ 85281 USA. (e-mail:madhurab@asu.edu; damoldov@asu.edu; rayyanar@asu.edu)
\par The materials presented in the papers are based upon work supported by the U.S. Department of Energy's Office of Energy Efficiency and Renewable Energy (EERE) under Solar Energy Technology Office Award Number DE-EE0008773.
}

}

\markboth{Journal of \LaTeX\ Class Files,~Vol.~14, No.~8, December~2023}%
{Shell \MakeLowercase{\textit{et al.}}: A Sample Article Using IEEEtran.cls for IEEE Journals}

\maketitle

\begin{abstract}
Volt-VAr control is a popular method for mitigating overvoltage violations caused by high penetration of distributed energy resources (DERs) in distribution feeders. An inherent limitation of volt-VAr control is that the reactive power ($Q$) absorbed/injected by the DER is determined based only on the terminal voltage, without considering the active power ($P$) generated by the DER. This leads to an inequitable burden of $Q$ support, in the sense that those DERs generating lower $P$, and hence contributing less to overvoltage issues, may be required to provide more than their share of $Q$ support.  The resulting PF of these DERs is required to vary over a wide range, which many current DERs do not support. A new control scheme, namely volt-PF control, is proposed here where the $Q$ support is inherently a function of both the voltage and $P$ from DERs,  which alleviates the above concerns while limiting the PF variation within a narrow range of 0.9 to 1. The proposed scheme is validated through extensive static and dynamic simulations on a real, large (8000+ nodes) feeder with very high penetration ($>$200\%) of DERs.The implementation of the new scheme in new and existing commercial hardware inverters is described.

\end{abstract}


\section{Introduction}
\IEEEPARstart{E}{nergy} generation in electrical grids is rapidly transitioning from non-renewable to renewable resources. Tackling climate change,  increasing regulations for fossil fuel, reducing costs for renewable energy, and favorable business models accelerate this transition to a greener grid. Compared with conventional energy resources, energy generated from variable energy resources (VER) such as wind and solar is expected to increase from 14\% in 2022 to 18\% in 2023 in the United States \cite{edi}. According to \cite{novel,snapshot}, a significant portion of the expected energy increase will be generated from distributed energy resources (DERs), mostly rooftop solar. 

\par Legacy distribution grids, designed for slower-response inductive loads, face challenges when integrating converter-interfaced generators (CIGs) with high intermittency \cite{harmonic_stab}, causing rapid voltage transients. Existing voltage regulation methods are ill-suited to manage these fast transients. Consequently, conventional voltage control and correction (VCC) devices struggle to respond swiftly to sudden changes in DERs due to fluctuations in irradiance \cite{distctrl}. This results in voltage violations across the distribution system, and also leads to degradation in the lifespan and performance of the VCC devices  \cite{slowoltc}. Additionally, the positioning of most VCC devices closer to the feederhead may hinder effective voltage corrections at the point of common coupling (PCC) for DERs.

\par Incorporating smart DERs for regulating the distribution grid voltage is popular because smart DERs not only monitor the PCC voltages right at the violations but are also capable of injecting/absorbing active and reactive power to/from the grid \cite{Vqimprove}-\cite{Rpowermng}. These can act as fast voltage regulators that can fill the gap between legacy voltage regulators and the fast dynamics of modern grids. CIGs can respond in a few microseconds to several milliseconds timeframe in contrast to the few seconds response times in VCCs \cite{definition}.

\par Inverter-based grid voltage regulation techniques are categorized into three groups: centralized, distributed, or a combination of both \cite{control_methods}. The centralized control achieves efficient and optimized grid voltage regulation with minimum impact on the distribution grid \cite{central1}-\cite{deeplearn}. While giving optimum results, centralized controls often suffer from higher complexity, high computational burden, and slower convergence compared to distributed controls.

\par Distributed control \cite{ancill}-\cite{Adjustable_robust} offers several key advantages over centralized control systems, such as smaller communication delays and not requiring complex control algorithms or dedicated communication architectures. Furthermore, it is impossible to determine the accurate load conditions, and the modeling of the feeder is not reliable due to topology changes regardless of the algorithm of the centralized control.   DER's ability to respond within several milliseconds ensures that grid dynamics, including sudden changes in load or intermittent renewable energy generation, are promptly accommodated \cite{DERdynamics}.  The lack of a dedicated communication medium in distributed control systems offers faster inverter response and supports inherent protection against unwarranted interferences and cyberattacks \cite{cyber}.


\par In many inverter-based voltage regulation schemes, the concept of fairness in allocating the burden of reactive power support among all the inverters to maintain feeder voltage regulation is often overlooked. However, from the consumer's perspective, ensuring fairness inappropriately sharing the grid support requirements is crucial for their support in strengthening the grid.  The dominant problem created by the high penetration of DERs in distribution feeders is the overvoltage violation, especially during high reverse power flow conditions created by excess DER generation. A fair voltage regulation scheme should ensure that a DER operating at lower $P$ generation should not be compelled to provide similar or higher $Q$ support compared to a DER operating at higher active power generation with the same PCC voltage ($V_{PCC}$).  Currently, popular voltage regulation schemes such as volt-VAr do not consider the operating active power of the DER and hence may demand high $Q$ support even from the DERs (of similar inverter VA ratings) that have low instantaneous power generation.  It may be noted that inverters of similar VA ratings on the AC side may still have a wide variation in the active power generation due to:

\begin{itemize}
    \item Different AC vs. DC power ratings of DERs and differences in the number of PV panels connected to inverters of the same VA rating. 
    \item Different orientations of PV panels.
    \item Partial shading of PV panels.
\end{itemize} 

\begin{table}[t]
\centering
\caption{Comparison of different control modes}
\begin{tabular}{|c|c|c|}
\hline
\multirow{2}{*}{\textbf{Control mode}} & \multirow{2}{*}{\textbf{Basis of $Q$ support}} & \multirow{2}{*}{\textbf{Comments}} \\
& &\\
\hline
 Volt-VAr & Based on $V_{PCC}$ & Does not ensure \\
 & only & fair $Q$ allocation\\
\hline
Constant PF & Based on $P$ only & Voltage regulation\\
& & not guaranteed \\
\hline
$P$ – $Q$ mode & Based on $P$ only & Voltage regulation\\
& & not guaranteed \\
\hline
Constant $Q$ & Independent of both & Voltage regulation \\
&$V_{PCC}$ and $P$ & not guaranteed \\
\hline
Voltage – $P$  & No $Q$ support; & Undesired power  \\
mode & $P$ curtailment  & curtailment\\
\hline
Proposed  & Based on both & Ensures fairness and\\
Volt-PF & $V_{PCC}$ and $P$ & voltage regulation \\
\hline
\end{tabular}
\label{comp_control}
\end{table}
\normalsize

\par The main reasons why requiring more than the appropriate share of reactive power support from DERs is unfair are as follows.
\begin{itemize}
    \item Absence of any monetary compensation to consumers for providing $Q$ support for voltage regulation.
    \item Shortening of the DERs lifespan due to extended operation at low PF \cite{lifetime}.
    \item Diminished operating efficiency of DERs due to higher currents associated with reactive power support.
    \item Reliability issues associated with the elevated stress on the electrolyte DC-link capacitors. These electrolytic capacitors are the weakest link in an inverter when it comes to reliability and durability, and their lifespan significantly degrades as the current through them increases.  Fig. \ref{dc_cur} shows the large increase in the instantaneous and RMS values of the current through the DC link capacitor in an inverter operating at a low PF.
\end{itemize} 

\par IEEE 1547-2018 \cite{IEEE1547} recommends several voltage control modes for feeder voltage management under normal operation.  These control modes are compared in Table I in terms of attaining fairness and the ability to regulate feeder voltage within prescribed limits.  Volt-VAr control can achieve voltage regulation, but since $P$ is not considered in determining $Q$ support, it does not lead to a fair allocation.  Both constant PF mode and active power-reactive power mode consider $P$ in determining $Q$ support.  However, since $V_{PCC}$ is not considered,  both of these schemes cannot guarantee voltage regulation.  Depending on the settings, these two schemes may under-compensate or over-compensate for voltage violations.  Constant $Q$ mode does not consider either $V$ or $P$ and hence cannot guarantee voltage regulation while also not ensuring fair $Q$ allocation.  Voltage-active power mode resorts to power curtailment, which is not desirable if the violations can be mitigated only by $Q$ support.  The limitations of the above control methods mainly arise from their single dimensional control, i.e., $Q$ support depends on only one (or less) measured variable.

\par The main objective of this paper is to propose a reactive power based voltage regulation (PF vs V) scheme for DERs that ensures voltage regulation as well as a fair allocation of $Q$ support based on the operating active power. In contrast to the popular volt-VAr control mode, where the reactive power is directly made a piecewise linear function of $V_{PCC}$ \cite{volt_var}, in the proposed method termed as Volt-PF mode, the operating PF of the DER is made a piecewise linear function of $V_{PCC}$. Such power factor control based on voltage, inherently ensures that the reactive power support demanded from DERs follows the operating active power of the DERs, leading to a fair allocation in addition to good voltage regulation. This type of voltage vs. PF control is already proposed for voltage regulation for weak distribution grids utilizing large-scale PV plants \cite{trans_vpf}. However, this paper proposed utilizing PF vs V curve implementation in distributed inverters in high penetration feeders and compared the performance with other IEEE 1547-2018 control modes.

\par The primary contributions of this paper can be summarized as follows:
\begin{figure}[t]
\centerline{\includegraphics[scale=0.32]{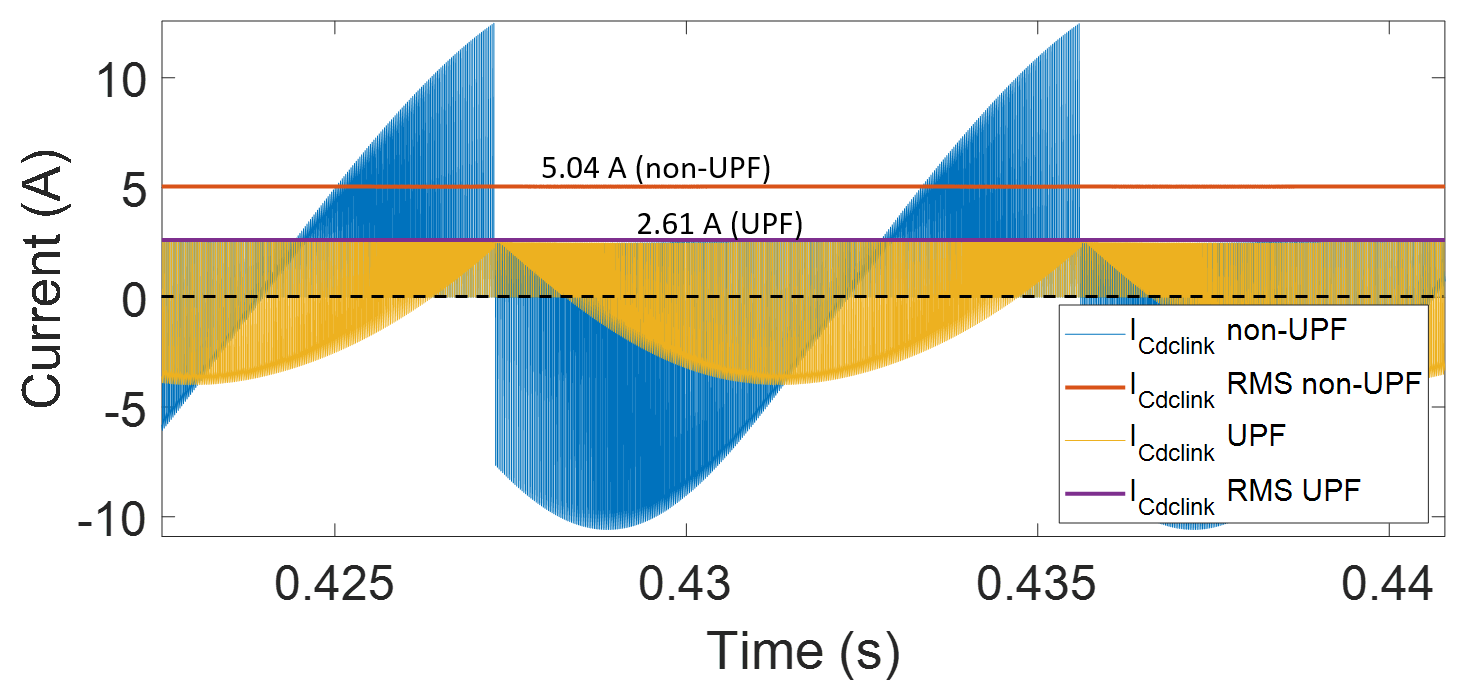}}
\caption{A comparison of the instantaneous and RMS values of the DC link capacitor current of an inverter operating at unity PF (1 kW) vs. operation at 0.5 PF (1 kW, 1.73 kVAr).}
\label{dc_cur}
\end{figure}

\begin{itemize}
  \item Introduction and complete development of the volt-PF control mode that features several advantages compared to volt-VAr and other IEEE 1547-2018 recommended control methods.
  \item Quantitative validation of the benefits of volt-PF (compared to volt-VAr under settings from multiple standards) in terms of significantly reduced range of variation in DER PF, reduced combined reactive power from DERs and transformer loading.  The validation has been performed through time-series and real-time simulations corresponding to a real feeder in Arizona with very high DER penetration.
  \item Implementation of the proposed volt-PF scheme in a custom-built hardware inverter as well as in a commercial inverter to demonstrate the ease of implementation.
  \item Variations of the proposed scheme with a similar objective of considering both voltage and active power in determining the $Q$ support from DERs. 
\end{itemize}

\section{Review of Volt-VAr Control}

\par
\begin{figure}[t]
\centerline{\includegraphics[scale=0.3]{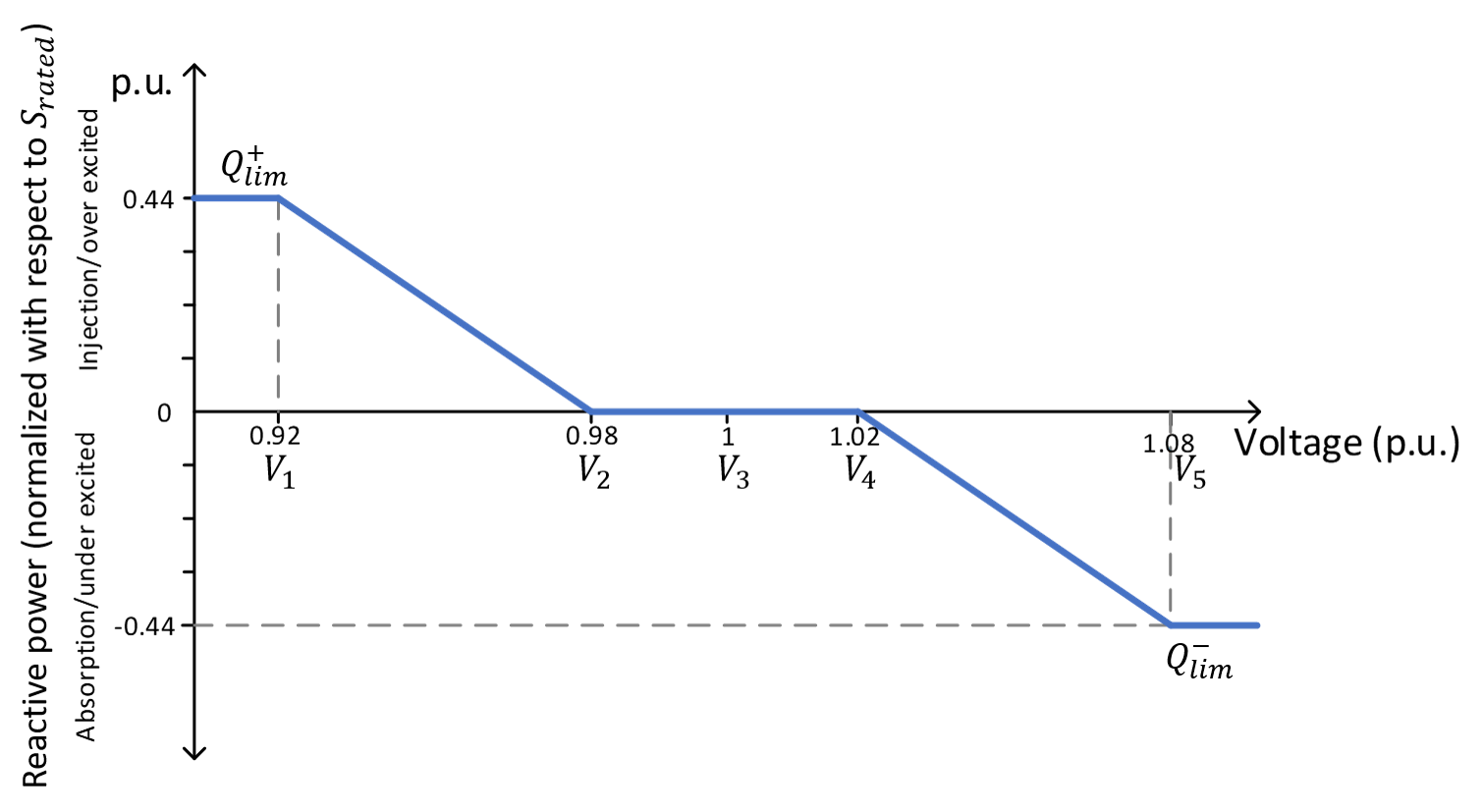}}
\caption{Default volt-VAr curve defined in IEEE Standard 1547-2018}
\label{fig1}
\end{figure}

 Fig. \ref{fig1}  shows the well-known volt-VAr characteristic featuring a piecewise linear relationship between the $V_{PCC}$, and the $Q_{support}$ from the DER.  The settings correspond to the default values given in IEEE standard 1547-2018.  When $V_{PCC}$ is in the dead zone between 0.98 p.u. and 1.02 p.u., $Q_{support}$ is zero. Above 1.02 p.u, $Q_{support}$ is absorbed, with its magnitude increasing linearly with voltage till  $Q_{support}$ = $Q_{limit}^-$ at $V_{PCC}$ = 1.08 p.u. At voltages below 0.98 p.u., $Q$ support is injected with the magnitude rising linearly as the voltage falls, till it reaches $Q_{limit}^+$ injection at $V_{PCC}$ = 0.92 p.u.  The volt-VAr expression, i.e, the $Q$ support from DER as a function of $V_{PCC}$ is shown in \eqref{eqnVV}, where $S_{rated}$ is the DER’s kVA rating.

{
\begin{equation}
\label{eqnVV}
Q_{support} =
    \begin{cases}
        0.44\;S_{rated} & V_{PCC} \leq V_1 \\
        (V_2- V_{PCC})\frac{0.44\;S_{rated}}{(V_2-V_1)}  &  V_1 < V_{PCC} < V_2\\
        0 &  V_2\leq V_{PCC} \leq V_4 \\
        -(V_{PCC}-V_4)\frac{0.44\;S_{rated}}{(V_5-V_4)} &  V_4< V_{PCC} < V_5 \\
        -0.44\;S_{rated} &  V_{PCC} \geq V_5
    \end{cases}
\end{equation}
}
\par As seen from (1) and Fig. \ref{fig1}, the $Q_{support}$ is independent of $P$ from the DER.  Hence, at low values of $P$, the PF of the DER can be quite low.  For example, when $P$ = 0.2 p.u., requirement of $Q_{support}$ = -0.44 $S_{rated}$ results in PF as low as 0.38.

\section{ Proposed Voltage - Power Factor (Volt-PF) Control Mode}

\par
\begin{figure}[t]
\centerline{\includegraphics[scale=0.28]{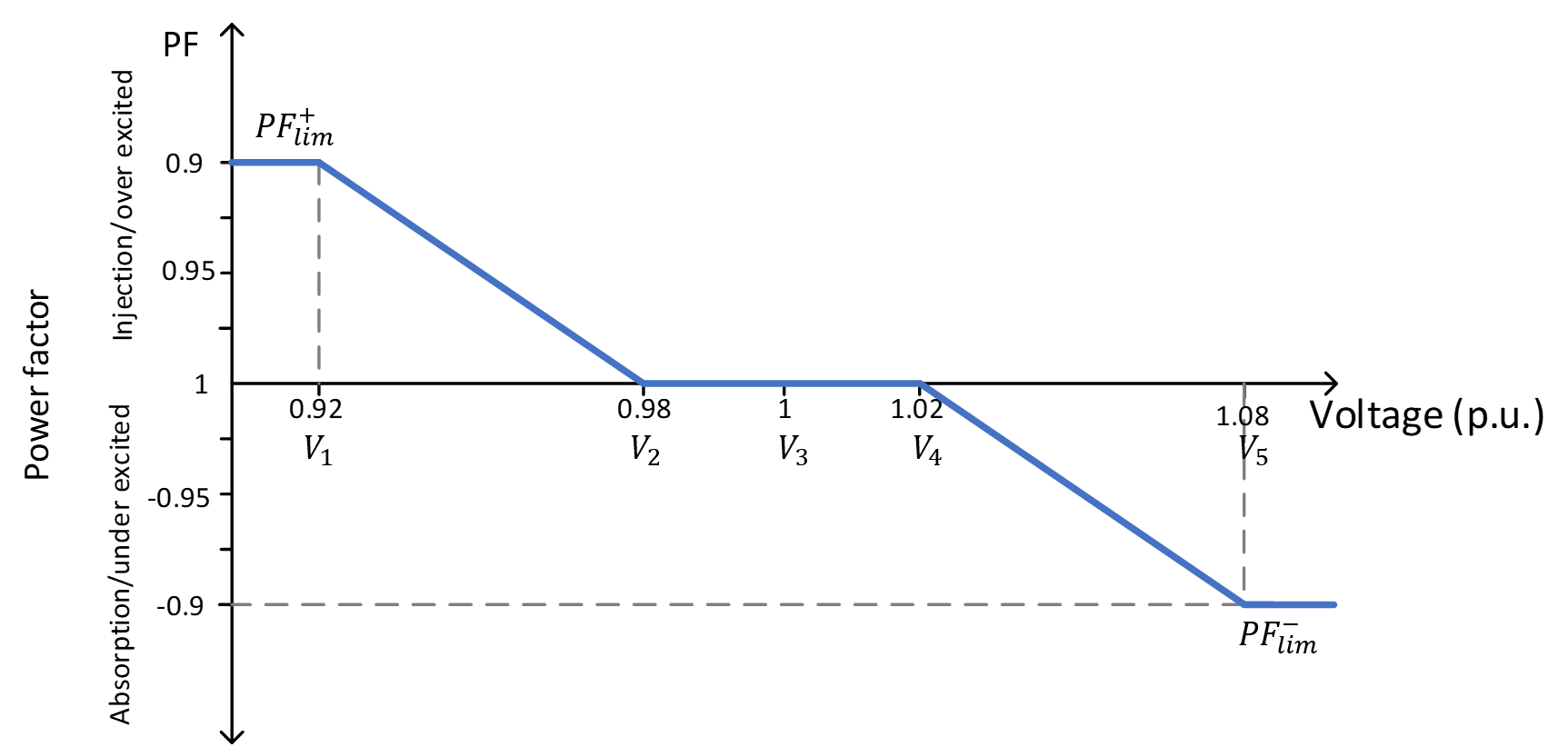}}
\caption{Proposed volt-PF characteristic with settings corresponding to those of default volt-VAR curve in 1547-2018}
\label{fig3}
\end{figure}

 The characteristic of the proposed volt-PF control is shown in Fig. \ref{fig3}, with the specific settings chosen to correspond to volt-VAr default settings.  As seen, the PF of DER rather than $Q$ is specified as a function of $V_{PCC}$ in the characteristic curve. Hence, the final $Q$ support from a DER depends on both $V_{PCC}$ and the DER active power (through PF specification).  When $V_{PCC}$ is in the dead zone between 0.98 p.u. and 1.02 p.u., the PF is unity. Above 1.02 p.u., PF is negative which corresponds to absorption of $Q$.  The PF falls linearly till it reaches $PF_{lim}^{-}$  at $V_{PCC}$ = 1.08 p.u. At voltages below 0.98 p.u.,  PF is positive corresponding to $Q$ injection.  The PF decreases linearly with $V_{PCC}$ till it reaches $PF_{lim}^{+}$ at $V_{PCC}$ = 0.92 p.u.  The proposed volt-PF characteristic, i.e, the PF of DER as a function of $V_{PCC}$, is shown in (2).

\par It may be noted that the set points $V_1$ to $V_5$ in Fig.\ref{fig3}  and (2)  correspond to those of IEEE 1547-2018 default volt-VAr settings and the $PF_{lim}^{+}$ and $PF_{lim}^{-}$ are chosen to obtain the $Q_{limit}^+$ and $Q_{limit}^-$ given in the volt-VAr default curve.  Each of these various settings can be modified based on the utility operating practices, feeder characteristics and DER penetration levels.  Section IV analyzes the performance of volt-PF control under different settings.

\par $PF_{lim}^+$ and $PF_{lim}^-$ indicate the PF saturation thresholds magnitudes for volt-PF mode operation under over-excitation and under-excitation conditions, respectively.

{
\begin{equation}
\label{eqnVPF}
PF  =
    \begin{cases}
         PF_{lim}^{+} & V_{PCC} \leq V_1 \\
        (1-(V_2- V_{PCC})\frac{(1-PF_{lim}^{+})}{(V_2-V_1)})  &  V_1< V_{PCC} < V_2\\
        1  &  V_2\leq V_{PCC} \leq V_4\\
        (-1+(V_{PCC}-V_4)\frac{(1 + PF_{lim}^{-})}{(V_5-V_4)})&  V_4< V_{PCC} < V_5 \\
        PF_{lim}^{-}  &  V_{PCC} \geq V_5
    \end{cases}
\end{equation}
}

\par Despite the linear relationship between voltage and PF, the relationship between voltage and reactive power is non-linear in the droop regions, the intervals from $V_1$ to $V_2$ and $V_4$ to $V_5$. The non-linear relationship of the reactive power in the over-excited droop region ($V_1$ to $V_2$) is governed by (\ref{eqn5}), while the under-excited droop region ($V_4$ to $V_5$) adheres to (\ref{eqn7}).

{
\begin{equation}
    \label{eqn4}
    Q(PF) = P\tan(\cos^{-1}(PF))
\end{equation}

If $V_{PCC} \leq V_1$,\\
\begin{equation}
    \label{eqn3}
     Q = P \tan(\cos^{-1}(PF_{lim}^+))
\end{equation}
\normalsize
If $V_1 < V_{PCC} < V_2$,\\

\begin{equation}
    \label{eqn5}
     Q = P \tan(\cos^{-1}(1 - \frac{(1- PF_{lim}^+)(V_2-V_{PCC})}{(V_2-V_1)})) 
\end{equation}
\normalsize
If $V_2< V_{PCC} < V_4$,
\begin{equation}
    \label{eqn6}
     Q = 0 
\end{equation}
\normalsize
If $V_4 < V_{PCC} < V_5$,
\begin{equation}
    \label{eqn7}
     Q = -P \tan(\cos^{-1}(1 - \frac{(1 + PF_{lim}^-)(V_{PCC}-V_4)}{(V_5-V_4)})) 
\end{equation}
\normalsize
If $V_5 \leq V_{pu}$,
\begin{equation}
    \label{eqn8}
     Q = P \tan(\cos^{-1}(PF_{lim}^-))
\end{equation}
}
\normalsize

\begin{figure}[t]
\centerline{\includegraphics[scale=0.23]{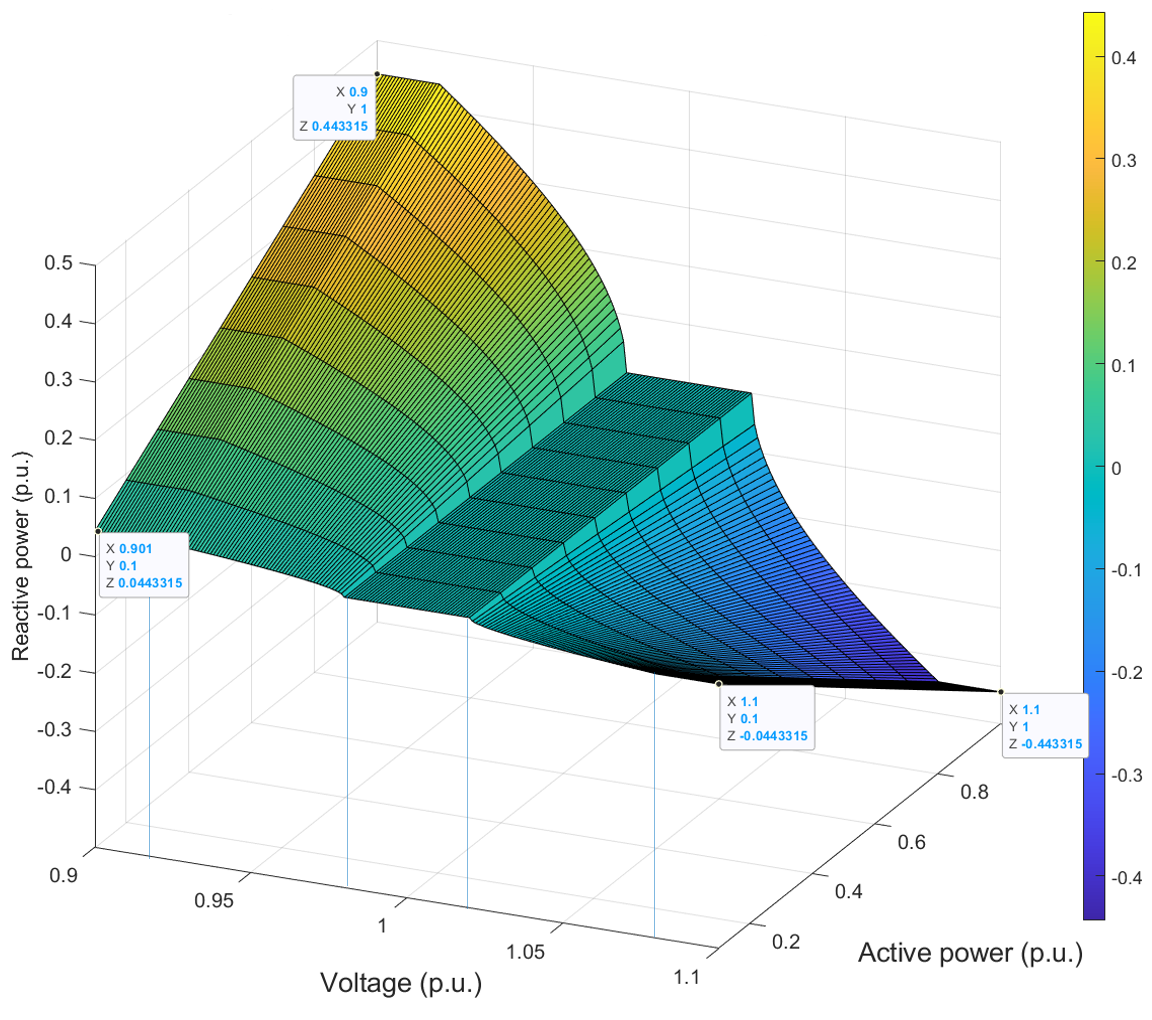}}
\caption{Volt-PF: Relationship between voltage, active power, and reactive power of an inverter.}
\label{3DVPQ}
\end{figure}

\begin{figure}[t]
\centerline{\includegraphics[scale=0.25]{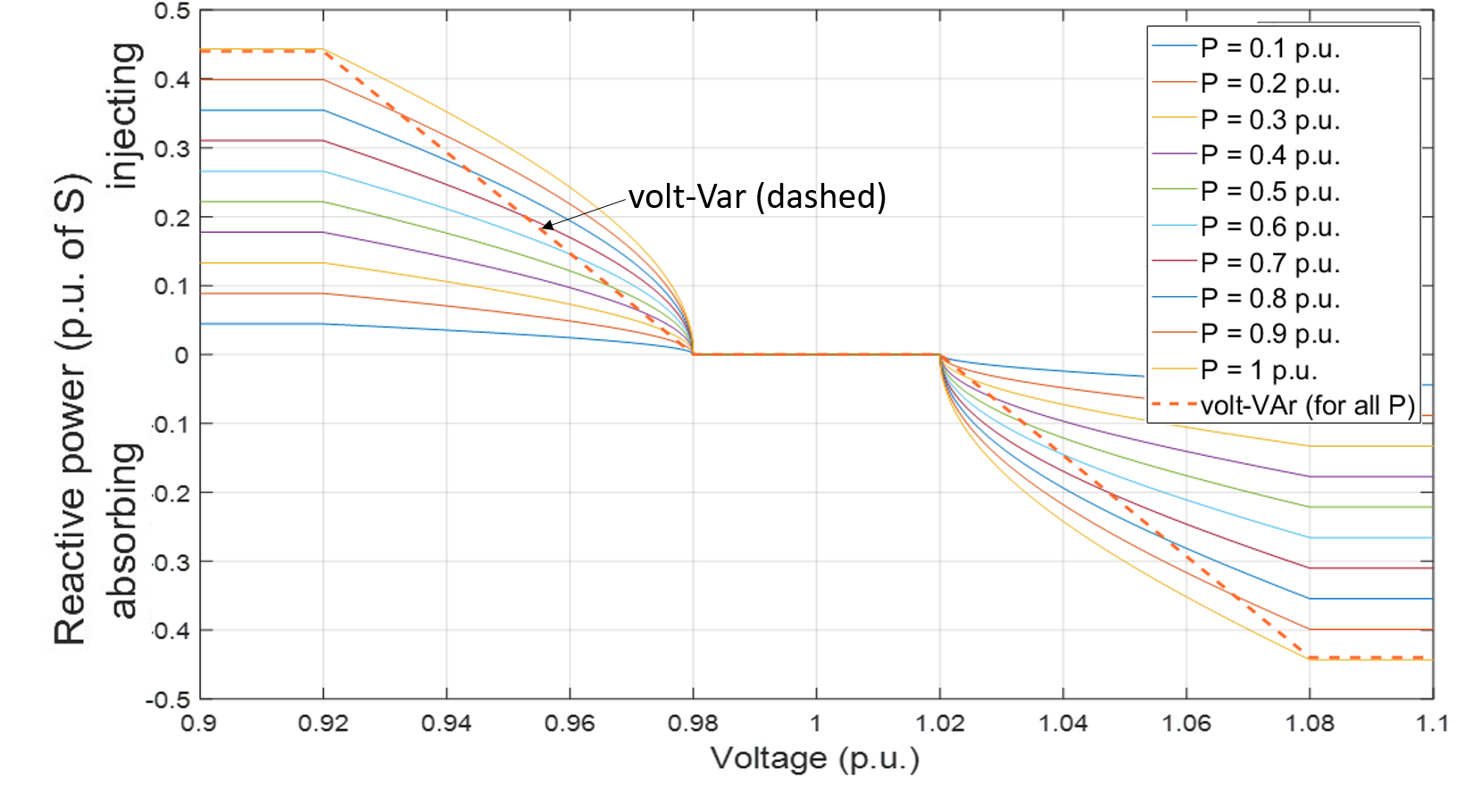}}
\caption{Volt-PF vs. volt-VAr:  Relationship between voltage and reactive power of an inverter under volt-PF mode (solid) compared with standard volt-VAr (dashed) for a range of inverter active power.}
\label{SliceVPQ}
\end{figure}

\par In volt-PF mode, the DER reactive power, as expressed by (\ref{eqn3}) to (\ref{eqn8}), relies on both $V_{PCC}$ and the active power generated by the DER. Fig. \ref{3DVPQ} shows a three-dimensional view of the dependence of DER $Q$ support on both  $V_{PCC}$ and the instantaneous DER active power when volt-PF control mode is used.  As seen, both for $Q$ absorption at high voltages and $Q$ injection at low voltages, the magnitude of $Q$ increases with the operating active power, $P$.  The maximum $Q$ support of +/- 0.44 $S_{rated}$ is provided only when the $P$ is at the full rated value, and for any other lower $P$, it is correspondingly lower, with a linear relation between $Q$ and $P$ at a given voltage.

\par Fig. \ref{SliceVPQ} shows a comparison of $Q$ vs. $V_{PCC}$ between volt-VAR and volt-PF control modes at different values of $P$.  First thing to note is that the volt-VAr has only a single curve for all $P$, where as in volt-PF, the characteristic is different for each $P$, with magnitude of $Q$ support increasing with $P$. Second is the non-linear relation between $Q$ and $V_{PCC}$ in the droop regions for volt-PF compared to a linear relation for volt-VAr.  The maximum $Q$ support for volt-VAr is always +/- 0.44 $S_{rated}$, while it is linearly reduced based on $P$ in the case of volt-PF, thus ensuring DERs with lower instantaneous active power are required to contribute only a smaller $Q$ support.  Volt-PF in general requires lower $Q$ support at a given $V_{PCC}$ except at regions close to the voltage deadband and high active power. 

\par Some key benefits of the volt-PF include,

\begin{itemize}
    \item Ensures voltage regulation along the feeder while enforcing DERs with higher P to provide higher $Q$ support. 
    \item Lower total $Q$ support needed compared to volt-Var (and other) schemes under similar settings.
    \item DER power factor is directly controlled (0.9 or better) while with volt-VAR, PF of DER is shown to be as low as 0.5 (some inverters cannot operate at this low PF).
    \item Lower $Q$ absorption from DERs operating at low $P$ also results in some savings in the feeder losses as well as some reduction in the loading of distribution transformers. 
    \item At the lower end of the high voltage condition (e.g., 1.02 to 1.03 p.u.) volt-PF requires a higher $Q$, resulting in DERs placed at lower voltage locations (near the feeder head, for example) also providing significant $Q$ contribution.
\end{itemize}

\par It may be noted that most DER inverters already support volt-VAr mode and constant PF mode, so volt-PF control can be effortlessly merged into new and existing smart inverters, as elaborated in Section V.

\section{Simulation Results From a High Penetration Feeder}

\subsection{Distribution Network}

\begin{figure}[h]
\centerline{\includegraphics[scale=0.34]{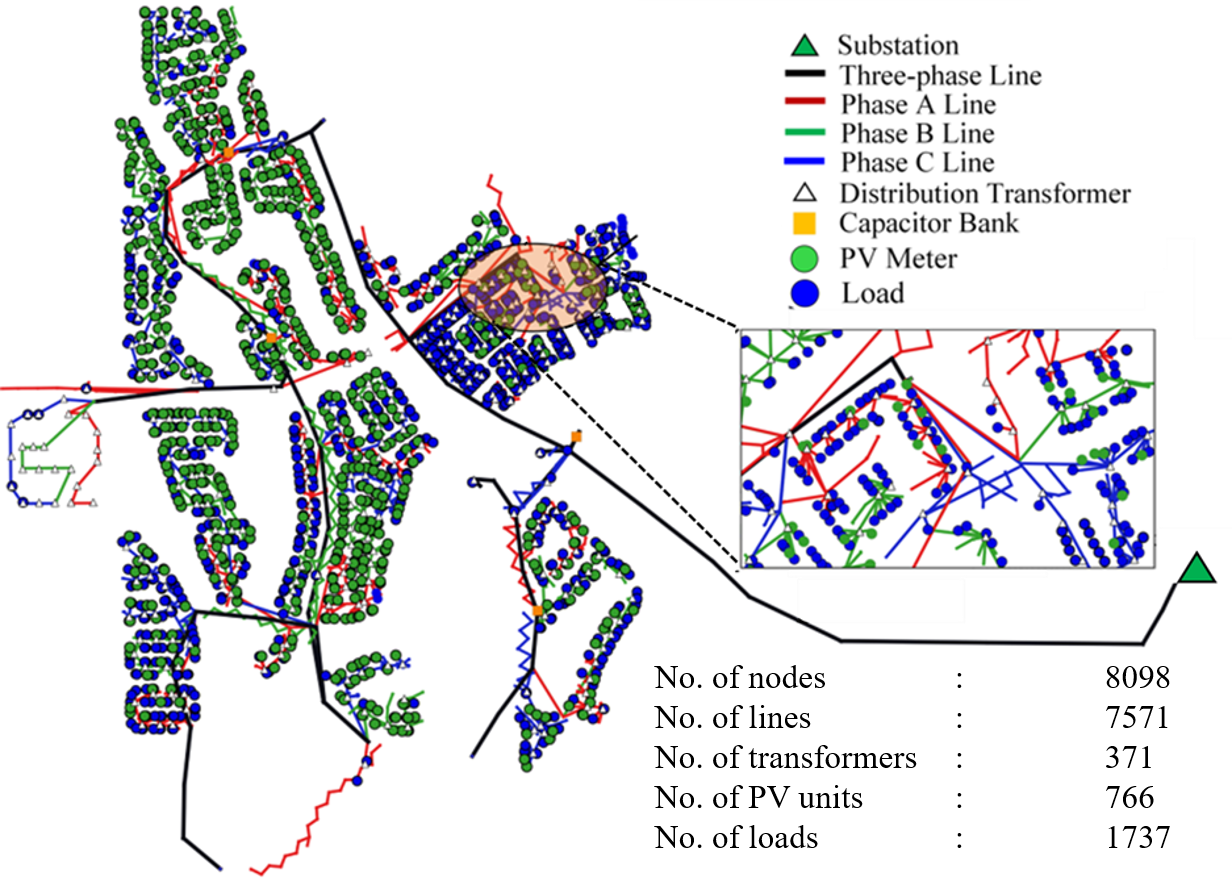}}
\caption{Distribution feeder selected to compare the effectiveness of the volt-PF mode vs. other existing control modes}
\label{feeder}
\end{figure}

\par The proposed volt-PF control method was evaluated through simulations on a feeder model based on an actual high-penetration distribution feeder \cite{motaz} in Arizona. Detailed information was extracted from a CYME network model obtained from the utility, and an accurate OpenDSS model was developed to analyze the feeder's behavior under various inverter operating conditions. The selected network, depicted in Fig. \ref{feeder}, experienced significant overvoltage violations during mid-spring due to lower loads and high PV generation.

\par At the peak of the violations, the network operated at 236.8\% instantaneous PV penetration. During this time, 29\% of the distributed nodes (2348 out of 8098) exceeded the voltage limit of 1.05 p.u. while 1732 consumer loads collectively drew 1.61 MW from the grid, and 766 DERs injected 3.8 MW of power into the grid. This resulted in an excess of 2.19 MW of power flowing back to the substation transformer, leading to substantial voltage violations throughout the grid, particularly at the consumer and distribution transformer nodes.

\par The analysis presented here examines the effectiveness of the volt-PF control in terms of voltage regulation, DER operating power factor, total feeder losses, and transformer loadings. The results from equivalent volt-VAr curves are also considered in the same scenarios alongside the proposed volt-PF curves to facilitate a comprehensive comparison. The evaluation uses both OpenDSS and ePHASORsim \cite{epsim} power system simulation tools, which provide accurate simulations of the real feeder in static and dynamic environments.

\par The simulation data for 24 hours comprises 24 snapshots, where each point represents an hour. Data aggregators of the utility service provider are utilized to collect profile information on individual loads, PV generators, and substation voltage. This dataset, consisting of 1732 active and reactive power load profiles, 766 PV profiles for active power injection, and the substation voltage profile, is then used in the feeder model for a realistic simulation during testing. 

\begin{figure}[t]
\centerline{\includegraphics[scale=0.45]{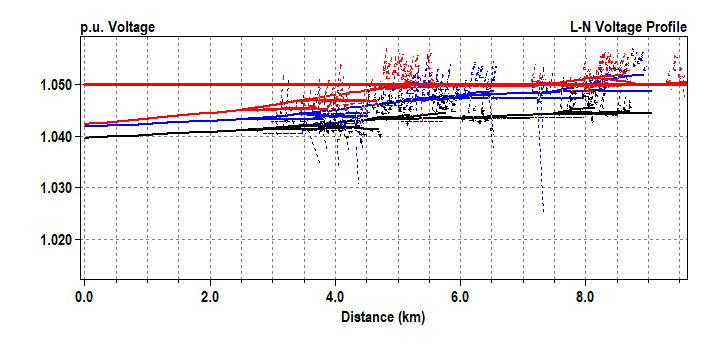}}
\caption{Voltage profile of the selected feeder plotted against the distance from the feederhead corresponding to UPF operation at the critical Hour 14. Black: Phase A, Red: Phase B, Blue: Phase C}
\label{vprof_base}
\end{figure}

\par Fig. \ref{vprof_base} shows the voltage profile along the feeder (voltage vs. distance from the substation) for the critical Hour 14. As seen, a large number of nodes have overvoltage violations (above 1.05 p.u.), especially in Phase B and Phase C.  Large reverse power flow results in voltage rising consistently from the substation towards the feeder end. 

\par
\begin{figure}[t]
\centerline{\includegraphics[scale=0.7]{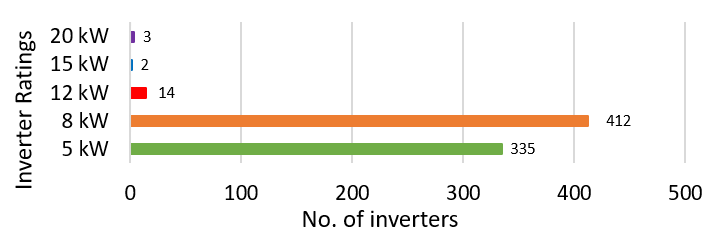}}
\caption{Number of inverters in each inverter rating category in the simulated feeder.}
\label{inv_rates}
\end{figure}

 The nameplate ratings of the inverters are not available from the utility.  Hence, based on the maximum power generation from each inverter over the entire dataset, the 766 inverters are grouped into five different ratings: 5 kW, 8 kW, 12kW, 15 kW, and 20 kW as shown in Fig. \ref{inv_rates}. The inverter kVA ratings, $S_{rated}$ are chosen such that there is no active power curtailment even when the inverter provides $Q$ support of 0.44 $S_{rated}$ at rated $P$. 

\par For dynamic simulations, the same 24-point dataset (March 15) is expanded into 1440-point profiles through MATLAB  using spline interpolation, thereby achieving a 1-minute resolution simulation. The network feeder model is converted from its static OpenDSS model to facilitate this dynamic simulation with each of the 766 inverters individually controllable. The simulation operates in real-time (with 1440 minute-profile condensed to 1440 seconds) at 2 ms time steps, providing smoother dynamic results for analyzing the feeder behavior under various inverter control schemes.

\subsection{Static Simulation Results}
\par The comparison is conducted between the volt-VAr and the volt-PF modes, highlighting the advantages of the proposed control mode using two distinct curve settings: 1. The IEEE 1547-2018 default curve, and 2. Hawaiian Electric SRD v1.1 curve.

\par For the static simulation in OpenDSS, all 766 inverters in the feeder are commanded to operate using both voltage regulation methods to support the grid. The assessment entails analyzing various parameters to evaluate the benefits and limitations of the volt-PF control scheme. The parameters considered for evaluation include total voltage violations, total feeder loss, total reactive power, inverter operating power factors, and the percentage loading of the transformers.

\begin{table}[t]
\centering
\caption{Feeder characteristics for volt-VAr and Volt-PF cases for IEEE 1547-2018 curve characteristics for Hour 14}
\begin{tabular}{|c|c|c|}
\hline
\multirow{2}{*}{Parameters} & \multirow{2}{*}{\textbf{Volt-VAr}} & \multirow{2}{*}{\textbf{Volt-PF}} \\
& & \\
\hline
Max voltage (p.u.) & 1.0446 & 1.0464 \\
\hline
Min voltage (p.u.) & 1.0118 & 1.0143 \\
\hline
Lowest DER power factor & 0.693 & 0.938\\
\hline
Total active power at feederhead (MW) & -2.02415 & -2.02638 \\
\hline
Total reactive power at feederhead (MVAr) & 1.1822 & 0.9806 \\
\hline
Total DER reactive power (MVAr) & -1.398 & -1.199 \\
\hline 
Total feeder active power losses (kW) & 88.15  & 85.92  \\
\hline
\end{tabular}
\label{comp_VPF_Std}
\end{table}
\normalsize
\subsubsection{The IEEE 1547-2018 default volt-VAr vs. equivalent volt-PF}

\begin{figure}[t]
\centerline{\includegraphics[scale=0.47]{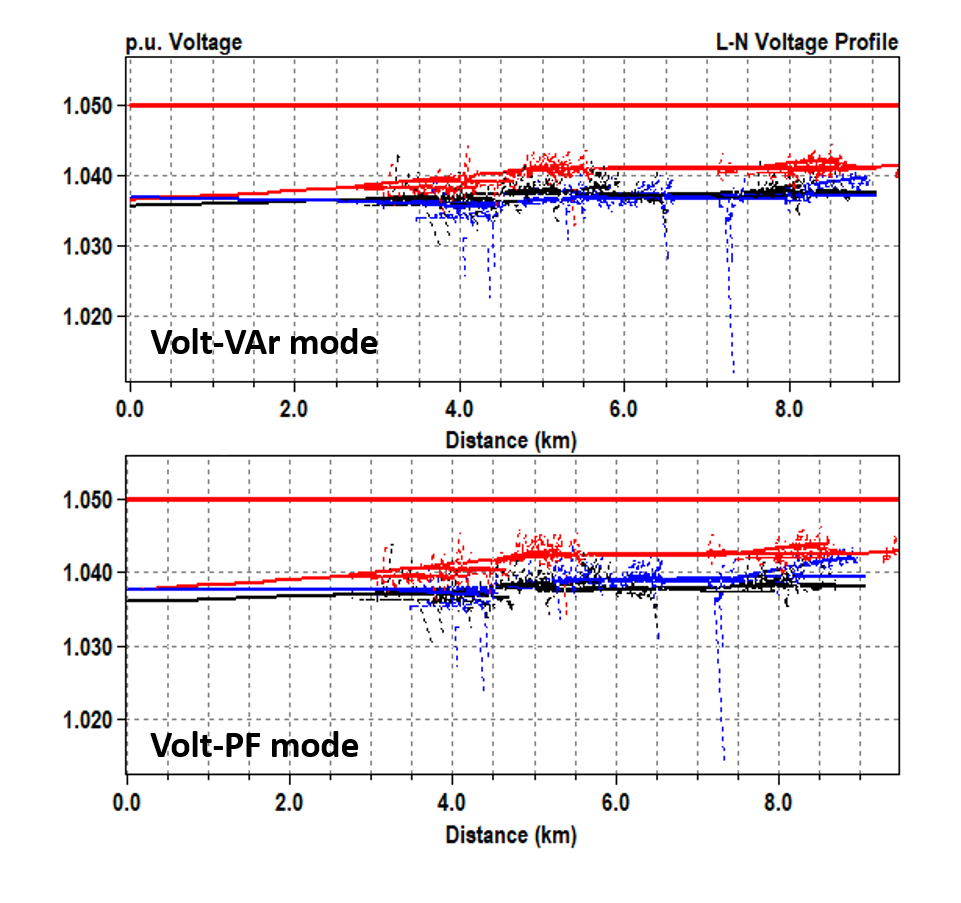}}
\caption{ Feeder voltage profiles at Hour 14 with the two voltage control schemes corresponding to IEEE 1547-2018 default. (a) 766 inverters operating in volt-VAr (top), (b) 766 inverters operating in volt-PF (bottom) }
\label{fig101}
\end{figure}

\par A comparable volt-PF curve setting, as depicted in Fig. \ref{fig3}, is replicated from the volt-VAr curve setting defined in the IEEE 1547-2018 standard shown in Fig. \ref{fig1}. The volt-PF curve is designed to maintain consistent voltage intervals for the over-excited, under-excited, dead band, and droop regions as in 1547-2018, while also adopting compatible high and low saturation limits (0.9 PF corresponds to 0.44 $S_{rated}$ at rated P). 

\begin{figure}[h]
\centerline{\includegraphics[scale=0.38]{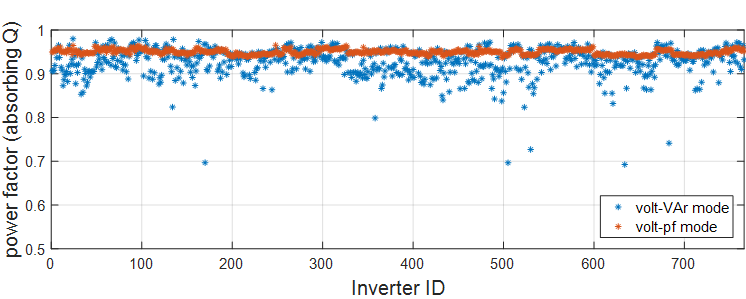}}
\caption{Inverter operating power factor at Hour 14 with volt-VAr (blue dots) and volt-PF (orange dots) corresponding to the IEEE1547-2018 default settings. }
\label{fig11}
\end{figure}

\par Fig. \ref{fig101} shows the feeder voltage profile with each of the two control schemes. As seen, both volt-VAr and volt-PF schemes completely eliminate the voltage violations. Although the voltage profiles seem nearly identical in both cases, the total reactive power from all the DERs combined is reduced by 17\% when using the volt-PF method compared to the volt-VAr mode. As a result, the active losses in the line are reduced, leading to a slight increase in reverse active power flow back into the substation for the volt-PF mode. Operational characteristics for the two scenarios are summarized in Table \ref{comp_VPF_Std} for Hour 14 simulation.

\begin{figure}[b]
\centerline{\includegraphics[scale=0.27]{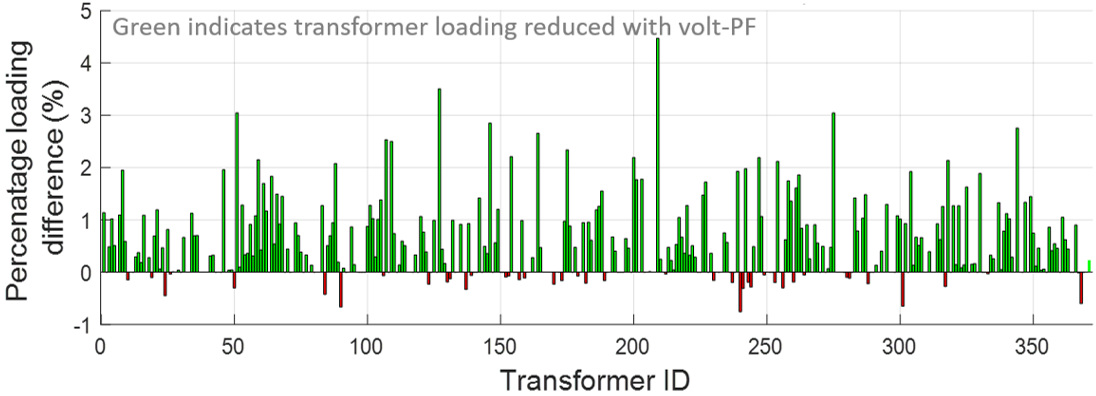}}
\caption{Transformer loading reduction for the volt-PF mode compared to the volt-VAr mode corresponding to IEEE 1547-2018 default settings. Very few transformers have higher loading with Volt-PF compared to Volt-VAr  and these are marked in red}
\label{fig13}
\end{figure}

\par Significant differences between the two methods are observed when examining their operation at the individual inverter and transformer levels. The volt-PF method is designed to inherently operate within a tight PF region at any operating $P$.  On the other hand, the volt-VAr control, which does not account for the inverter's operating $P$, results in a wide variation in the PF, between 0.978  and 0.693, compared to between 0.965 and 0.938 for volt-PF, as shown in Fig.\ref{fig11}. Inverters that operate at lower $P$ and are located at high voltage nodes absorb higher $Q$, leading to low PF.  

\par The difference in the reactive power absorbed also impacts the loading of the distribution transformers to which the DERs are connected.  Most loads absorb Q (inductive loads), and hence, the increased Q absorption by DERs for grid support contributes to increasing the transformer loading regardless of whether the active power flow direction is positive or negative.  As seen in Fig. \ref{fig13}, the volt-PF results in around 1\%-4\% reduction in the kVA loading of most of the transformers that have DERs.  Only in a few transformers, the volt-PF scheme results in higher loading compared to volt-VAr.

{
\begin{table}[h]
\centering
\caption{Feeder characteristics for volt-VAr and Volt-PF cases for  curve Hawaiian Electric SRD v1.1 characteristics for Hour 14}
\begin{tabular}{|c|c|c|}
\hline
\multirow{2}{*}{Parameters} & \multirow{2}{*}{\textbf{Volt-VAr}} & \multirow{2}{*}{\textbf{Volt-PF}} \\
& & \\
\hline
Max voltage (p.u.) & 1.0409 & 1.0445 \\
\hline
Min voltage (p.u.) & 1.0061 & 1.0121 \\
\hline
Lowest DER power factor & 0.513 & 0.909\\
\hline
Total active power at feederhead (MW) & -2.0172 & -2.0241 \\
\hline
Total feederhead reactive power (MVAr) & 1.6695 & 1.1924 \\
\hline 
Total DER reactive power (MVAr) & -1.875 & -1.408 \\
\hline
Total active loses (kW) & 95.1  & 88.2  \\
\hline
\end{tabular}
\label{comp_VPF_Hawaii}
\end{table}
}
\subsubsection{Comparison of volt-PF and volt-VAr corresponding to settings in Hawaiian Electric SRD v1.1}

\begin{figure}[b]
\centerline{\includegraphics[scale=0.45]{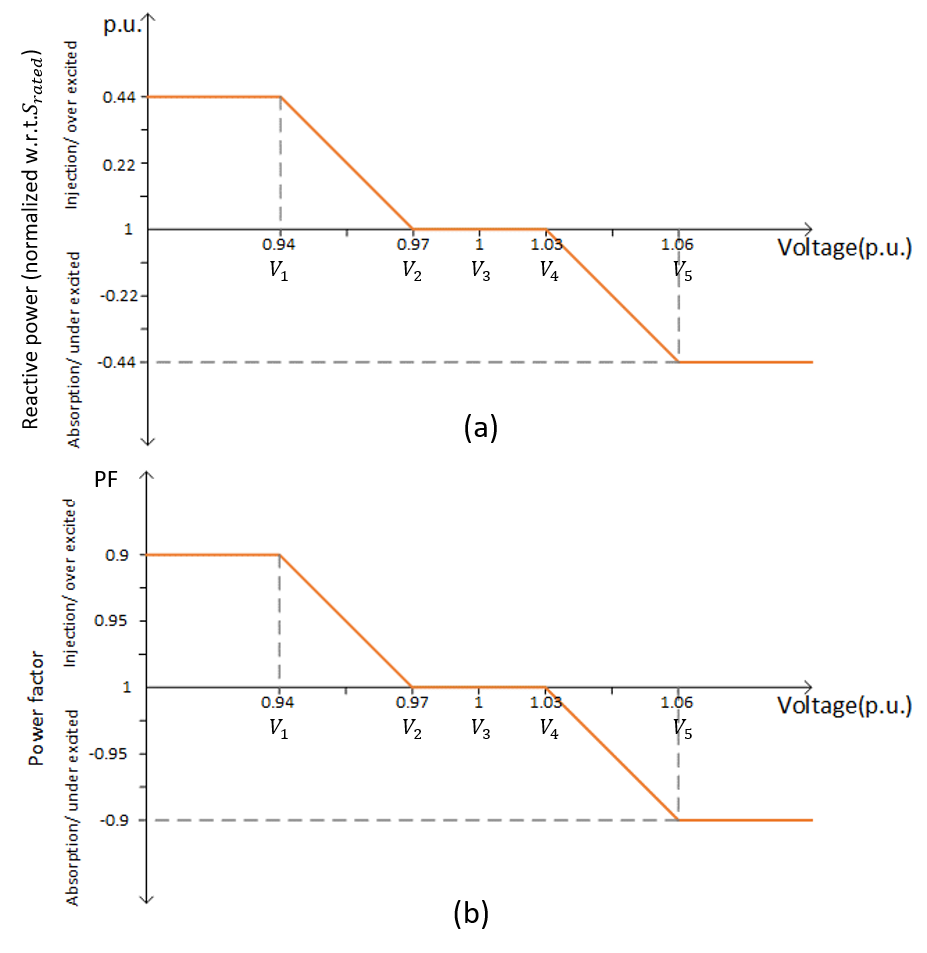}}
\caption{Volt-VAr curve (top) vs equivalent volt-PF curve (bottom) based on the Hawaiian Electric SRD v1.1}
\label{fig14}
\end{figure}

\par Fig. \ref{fig14}a shows the default volt-VAr curves and settings from Hawaiian Electric SRD v1.1 and Fig.\ref{fig14}b shows the corresponding curve and settings for the proposed volt-PF control mode.

\begin{figure}[htb]
\centerline{\includegraphics[scale=0.35]{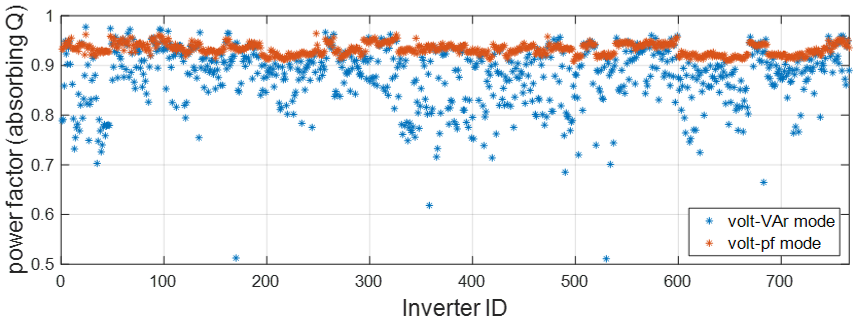}}
\caption{Inverter operating power factor at Hour 14 with the two voltage control schemes corresponding to Hawaiian Electric SRD v1.1 default settings. (a). 766 inverters operating in volt-VAr (Blue), (b). 766 inverters operating in volt-PF (Orange) }
\label{fig16}
\end{figure}

\begin{figure}[htb]
\centerline{\includegraphics[scale=0.25]{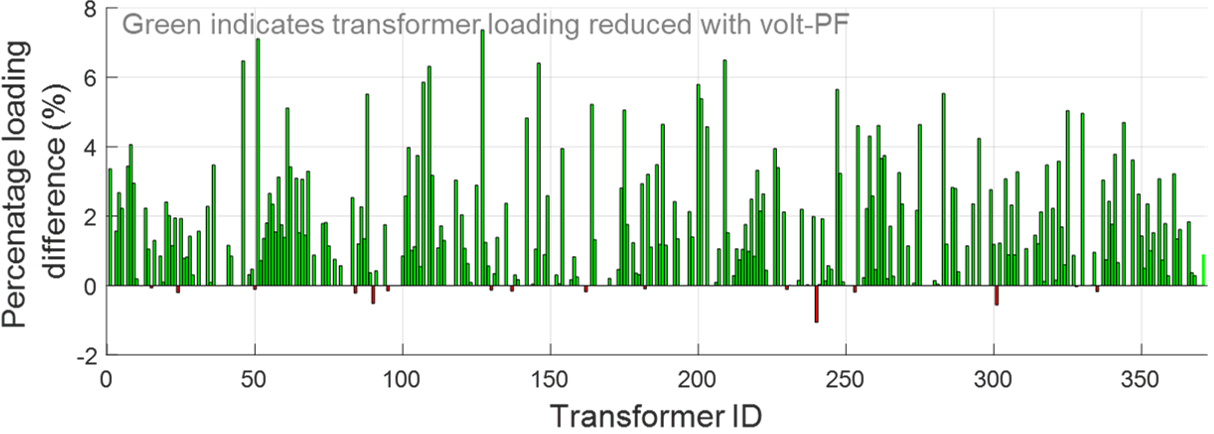}}
\caption{Transformer loading reduction for the volt-PF mode compared to the volt-VAr mode corresponding to Hawaiian Electric SRD v1.1 settings. Very few transformers have higher loading with Volt-PF compared to Volt-VAr, and these are marked in red}
\label{fig18}
\end{figure}

\par In both instances, the voltage regulation schemes based on Hawaiian Electric SRD v1.1 curve settings prove effective in mitigating all voltage violations when all inverters are instructed to switch to the designated grid support mode, be it volt-VAr or volt-PF. As summarized in Table \ref{comp_VPF_Hawaii}, the total DER reactive power when volt-PF is employed is 28.5\% lower compared to volt-VAr, leading to about 7.2\% reduction in the active power losses with volt-PF mode. 

\par The inverter's operating PF also exhibits a comparable trend to the IEEE 1547-2018 standard curve setting.  As presented in Fig.\ref{fig16}, the volt-PF mode enforces a more stringent control on the power factor with all DERs operating between 0.964 and 0.909 PF.  In comparison to volt-VAr, the PF varies widely from 0.979 to as low as 0.513.
Compared to the volt-VAr mode, the individual distribution transformers also experience reduced loading for the most. As shown in Fig.\ref{fig18}, the loading decreases by up to 7\% in certain cases, with a typical reduction of around 3-5\% for most.

\subsection{Dynamic Simulation Results}

\par The performance of various DER control modes in terms of voltage regulation and required $Q$ support is also compared through dynamic simulations.  For this purpose, an Opal-RT ePHASORsim model of the feeder and DERs is generated from the OpenDSS model.  The details of the conversion process are explained in detail in \cite{PVSC}.  Ten of the DER inverters are modeled in full detail in the state space domain and include the LCL filter, phase lock loop (PLL) for grid synchronization and various controllers. To minimize the computation burden, the remaining inverters are modeled as controlled current sources capable of supporting any of the control modes, including the proposed volt-PF mode.  Before conducting extensive dynamic simulations, the accuracy of the ePHASORsim model is validated by comparing it with the OpenDSS static model.  The maximum voltage magnitude error between the two models is 0.27\%, with 99\% of the errors less than 0.1\%.   
  
\par
\begin{figure}[htb]
\centerline{\includegraphics[scale=0.25]{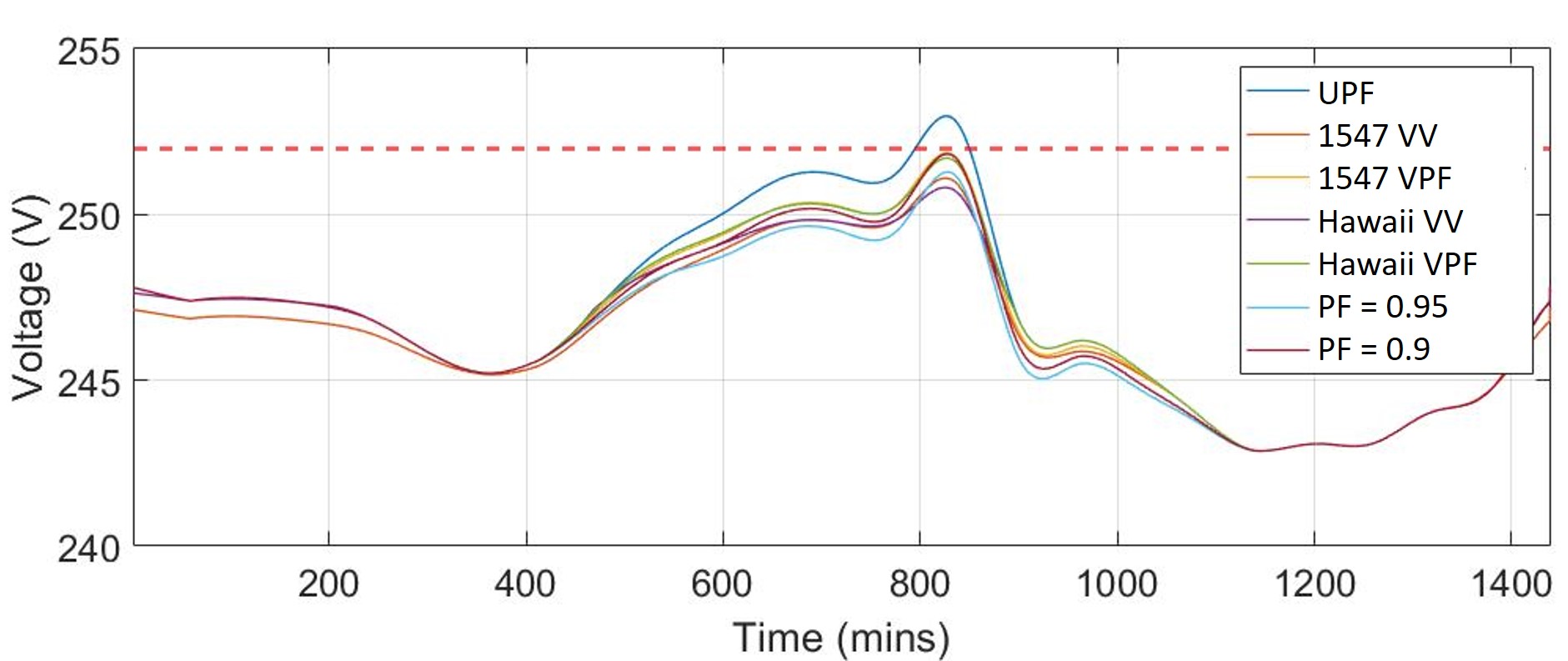}}
\caption{Voltage of a selected inverter in dynamic simulation for different voltage regulations 1. Unity PF 2. Volt-VAr IEEE 1547-2018 curve 3. Volt-PF curve based on 1547 curve 4. Volt-VAr Hawaiian curve 5. Volt-PF curve based on Hawaiian 6. Constant PF operation 0.95 7. Constant PF operation 0.9}
\label{fig191}
\end{figure}

\begin{figure}[htb]
\centerline{\includegraphics[scale=0.26]{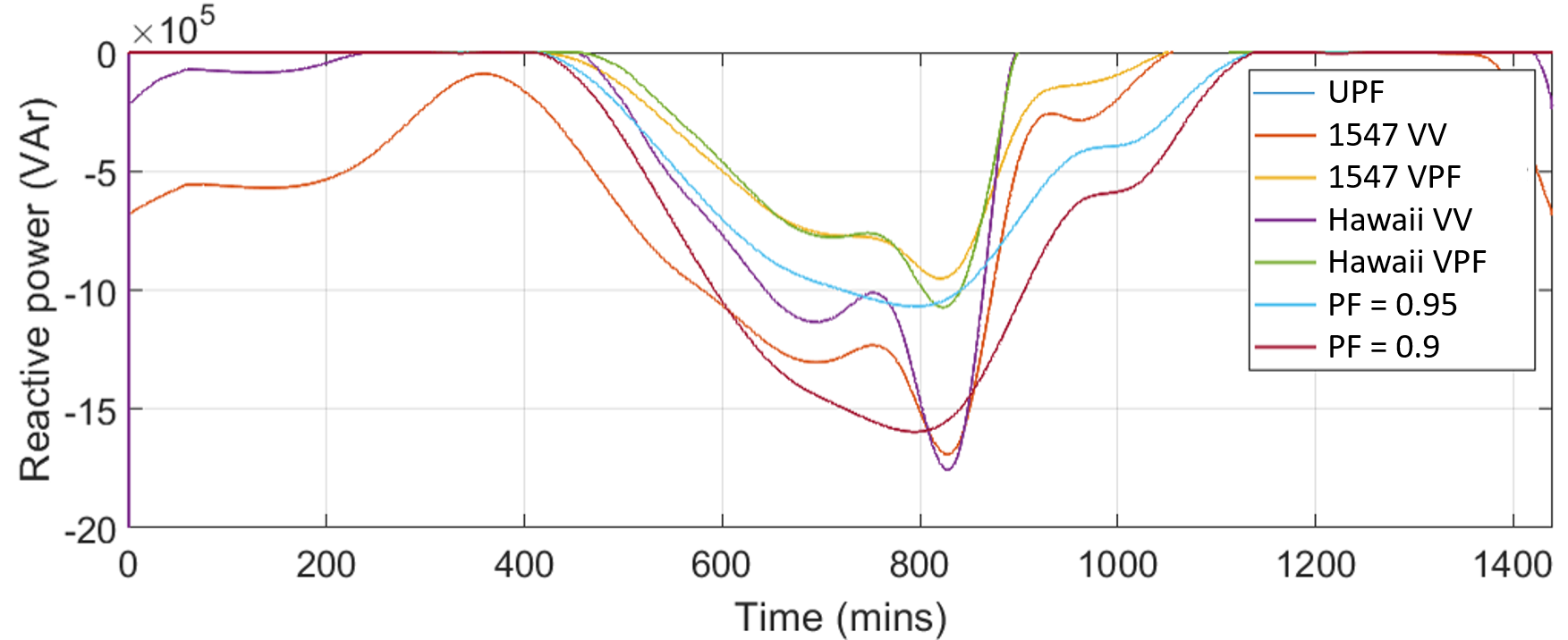}}
\caption{Total reactive power utilized in voltage regulation for different modes: 1. Unity PF 2. Volt-VAr IEEE 1547-2018 curve 3. Volt-PF curve based on 1547 curve 4. Volt-VAr Hawaiian curve 5. Volt-PF curve based on Hawaiian 6. Constant PF operation 0.95 7. Constant PF operation 0.9}
\label{fig201}
\end{figure}

 The dynamic model is programmed to operate with various control modes for feeder voltage management including the five control modes recommended in IEEE 1547-2018 and the proposed volt-PF.  Different settings for these control modes were also programmed including default settings from IEEE 1547-2018 and Hawaiian Electric SRD v1.1.  Fig. \ref{fig191} illustrates the variation in the terminal voltage of a selected DER inverter over a 24-hour cycle, considering different grid support modes. In the base case, where the feeder operates at unity PF, over-voltage violations occur from Hour 10 to Hour 14. Each of the grid support control modes studied, including volt-VAr, constant PF, and volt-PF under different settings, are able to mitigate the overvoltage violations under all hours for the selected inverter. However, the total reactive power from the DER inverters varies significantly across the different modes. A control mode that can maintain the maximum voltage of the feeder within limits with minimal reactive power usage can be considered more efficient than others. As depicted in Fig. \ref{fig201}, the two volt-PF curves lead to significantly lower utilization of reactive power during peak PV penetration hours while effectively regulating the grid voltage. When compared with the volt-VAr results, the volt-PF modes use 43.8\% (1690.9 kVAr vs. 949.8 kVAr) and 38.8\% (1753.9 kVAr vs. 1071.8 kVAr) less peak reactive power for IEEE 1547-2018 standard curve setting and Hawaiian Electric SRD v1.1 curve settings respectively.

\section{Hardware results}
\par The proposed volt-PF scheme is implemented in a custom-built hardware solar PV inverter. The implementation of the volt-PF code follows a simple algorithm shown in Algorithm \ref{vPF_alg}. The PF command from the selected volt-PF curve together with the voltage measurement at the point of interconnection and known active power command (such as from MPPT control loop) are used to generate the reactive power reference ($Q_{ref}$). A four-quadrant grid simulator with programmable voltage magnitude is used to verify the performance of the volt-PF implementation.


{
\begin{algorithm}[t]
\caption{The volt-PF implementation}\label{vPF_alg}

1.\hspace{0.1cm}  \textbf{If} (inverter control mode = volt-PF)\\
2.\hspace{1 cm}  \textbf{If} ($V_{PCC} \leq V_1$) ; $PF = PF_{lim}^+$\\
3.\hspace{1 cm}  \textbf{else if} ($V_1 \leq V_{PCC} \leq V_2$) ;\\
{
\begin{equation*}
PF = (1-(V_2- V_{PCC})\frac{(1-PF_{lim}^+)}{(V_2-V_1)})\\
\end{equation*}
}
4.\hspace{1 cm}  \textbf{else if} ($V_4 \; < \;V_{PCC} \; < \; V_2$) ; $PF = 1$\\

5.\hspace{1 cm}  \textbf{else if} ($V_3\; < \; V_{PCC} \; < \; V_4$) ;\\
{
\begin{equation*}
PF = (-1+(V_{PCC}-V_4)\frac{(1+PF_{lim}^-)}{(V_5-V_4)})\\
\end{equation*}
}
6.\hspace{1 cm}  \textbf{else if} ($V_5 \leq V_{PCC} $) ; $PF = PF_{lim}^-$\\
7.\hspace{0.1cm} \textbf{End}

\end{algorithm}
}

\par The PF command from the algorithm is converted into two parameters: active power ($P_{ref}$) and reactive power references ($Q_{ref}$), which are then fed into the control code. Different curve settings can be stored in or communicated to the inverter through the communication channel, similar to volt-VAr curve modifications.


\begin{figure}[t]
\centerline{\includegraphics[scale=0.3]{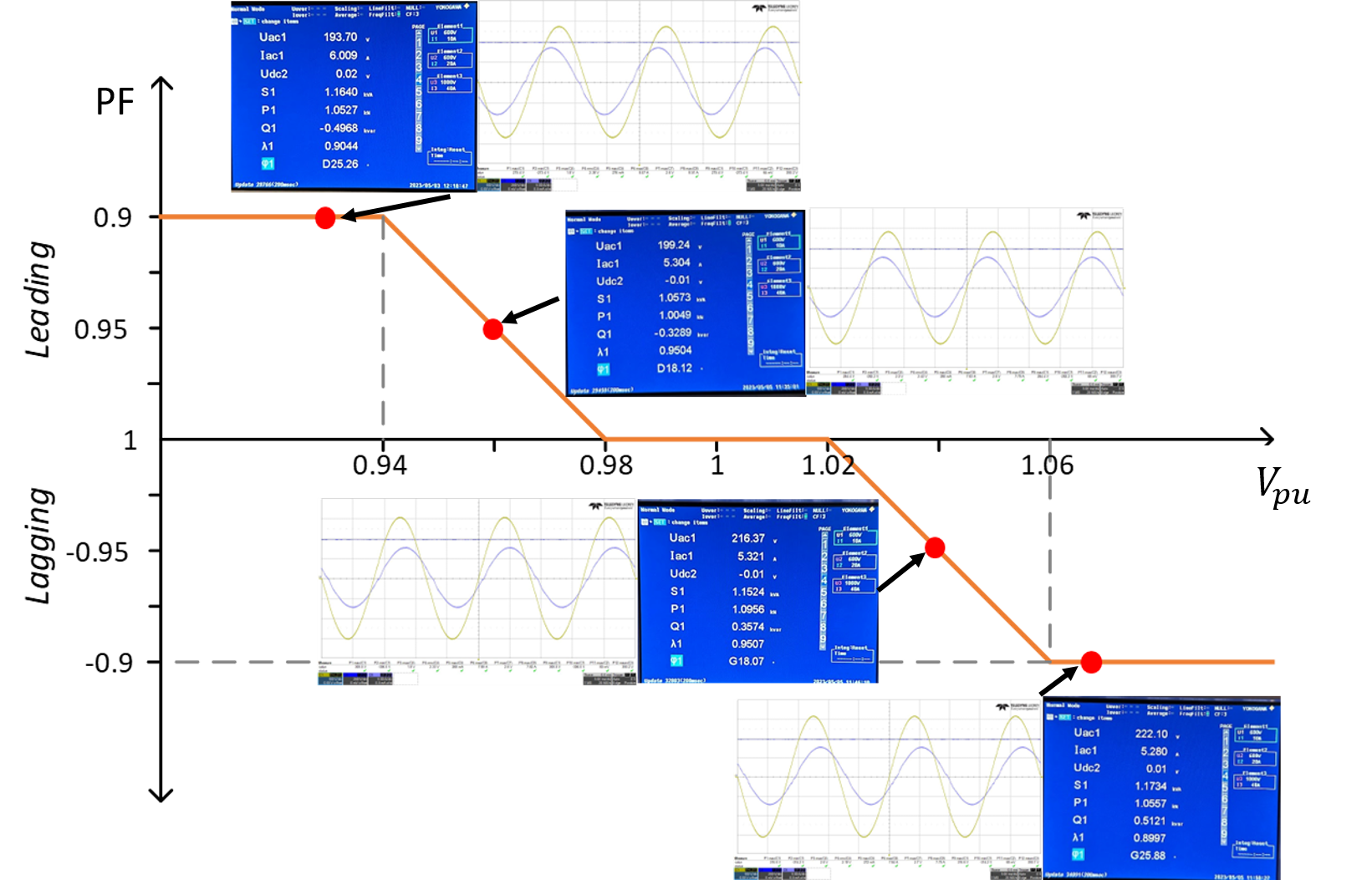}}
\caption{Hardware results for the custom-built inverter operating in the volt-PF mode at different voltage intervals}
\label{fig19}
\end{figure}

\par Fig. \ref{fig19} shows the hardware results from the custom-built inverter corresponding to the volt-PF operation with settings similar to the default volt-VAr settings in IEEE 1547-2018 curve. The inverter is set to operate at 1 kW, with injecting and absorbing PF limits set to 0.9. The respective oscilloscope waveforms illustrate the inverter's operation, and the power analyzer results further confirm the accurate operation of the inverter under the volt-PF mode.

\par Implementing volt-PF mode in commercial inverters as a new control mode is also relatively simple since most commercial inverters support the IEEE 1547-2018 recommended constant PF mode as a standard feature.  This feature can be leveraged to make the operating PF a function of the terminal voltage as derived from the volt-PF curve.  Since the inverter terminal voltage is essential for various other control, synchronization, and protection functions, volt-PF does not require any additional sensors or other hardware.

\par Since the control code of commercial inverters cannot be directly changed by the user, we make use of the Modbus communication line in commercial inverters designed for utilities to issue commands for control mode selection and adjust the curve settings.  In our project, the team has developed an edge intelligent device (EID) for providing complete solar situational awareness to the utilities and to enable real-time, coordinated DER control \cite{yunpeng}.  By employing the EID equipped with computational capabilities, it becomes possible to control the inverter through the Modbus communication channel. Within the EID, the volt-PF control scheme is housed, and it utilizes Algorithm \ref{vPF_alg} to generate precise PF commands corresponding to the operating voltage. Subsequently, these calculated PF commands are sent back to the commercial inverters, allowing them to adhere to the volt-PF curve seamlessly.


\begin{figure}[t]
\centerline{\includegraphics[scale=0.3]{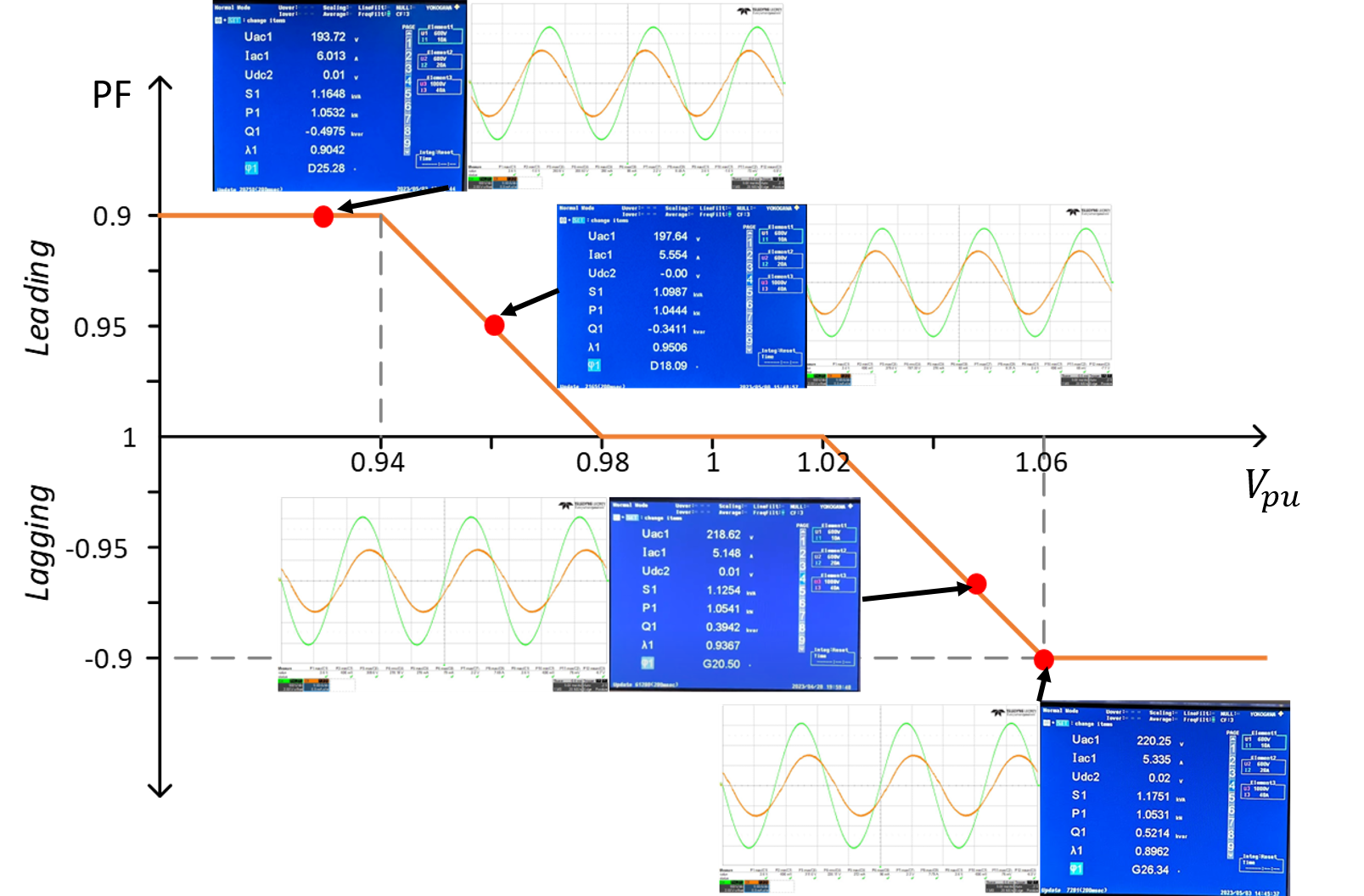}}
\caption{Hardware results for the commercial inverter commanded from an EID operating in the volt-PF mode at different voltage intervals}
\label{fig20}
\end{figure}

\par Fig. \ref{fig20} illustrates a commercial inverter from SMA that effectively follows the volt-PF curve through the mentioned EID control loop. The volt-PF curve is integrated into the EID code, allowing the EID to calculate the current PF command. This command is then sent to the inverter via the  Modbus channel. The added communication loop introduces a slight delay to the control, typically under 1 second, which has no significant impact on the voltage regulation capability.

\section{Modified volt-PF curve with linear V-Q relationship}

\par
\begin{figure}[t]
\centerline{\includegraphics[scale=0.3]{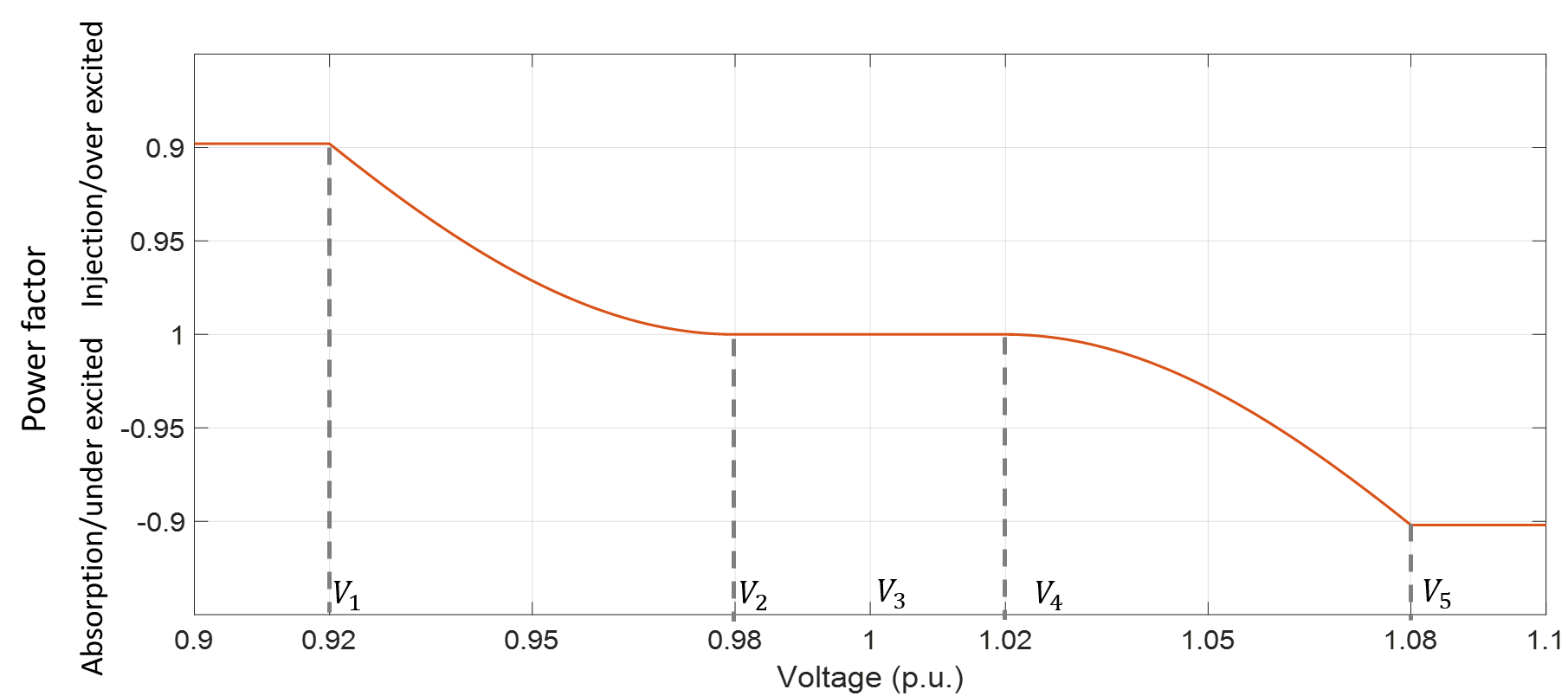}}
\caption{PF vs. $V_{PCC}$ for  the proposed volt-PF ver. 2 control scheme.}
\label{fig21}
\end{figure}

 In the volt-PF curves introduced in previous sections (Fig. \ref{fig3} and Fig. \ref{fig14} for example), the relation between the voltage and reactive power is governed by piecewise non-linear equations, specifically (\ref{eqn5}) and (\ref{eqn7}) which apply during the interval [$V_1$, $V_2$] and [$V_4$, $V_5$] respectively.  Initial feedback from some of the stakeholders (such as working groups for DER interconnection standards) suggests that a linear relation between $V$ and $Q$ in the above intervals may be preferred.  Hence, for situations where the linear $V$-$Q$ relation is preferred, we propose a modified volt-PF curve - volt-PF ver. 2 that has a piecewise linear relation between $V$ and $Q$ while still maintaining the fair allocation of reactive power support and limited variation of PF.   Incorporating linear relation between $V$ and $Q$, makes the relation between $V$ and power factor non-linear.  The governing equations for volt-PF ver. 2 are presented in \eqref{eqnVPF2} which is very similar to the volt-VAR equation in (1) except for the scaling factor $P/P_{rated}$.  The defining PF vs. voltage curve for the new scheme is shown in Fig. \ref{fig21}.  For the settings shown in Fig. \ref{fig21}, the operating PF of the DER always remains within 0.9 to 1 (lagging or leading).  

{\small
\begin{equation}
\label{eqnVPF2}
Q_{VPF2} =\frac{P}{P_{rated}}
    \begin{cases}
        0.44 S_{rated} & V_{PCC} \leq V_1 \\
        (V_2- V_{PCC})\frac{0.44 S_{rated}}{(V_2-V_1)}  &  V_1 < V_{PCC} < V_2\\
        0 &  V_2\leq V_{PCC} \leq V_4 \\
        (V_4-V_{PCC})\frac{0.44 S_{rated}}{(V_5-V_4)}) &  V_4< V_{PCC} < V_5 \\
        -0.44 S_{rated} &  V_{PCC} \geq V_5
    \end{cases}
\end{equation}
}

\normalsize

\par
\begin{figure}[t]
\centerline{\includegraphics[scale=0.28]{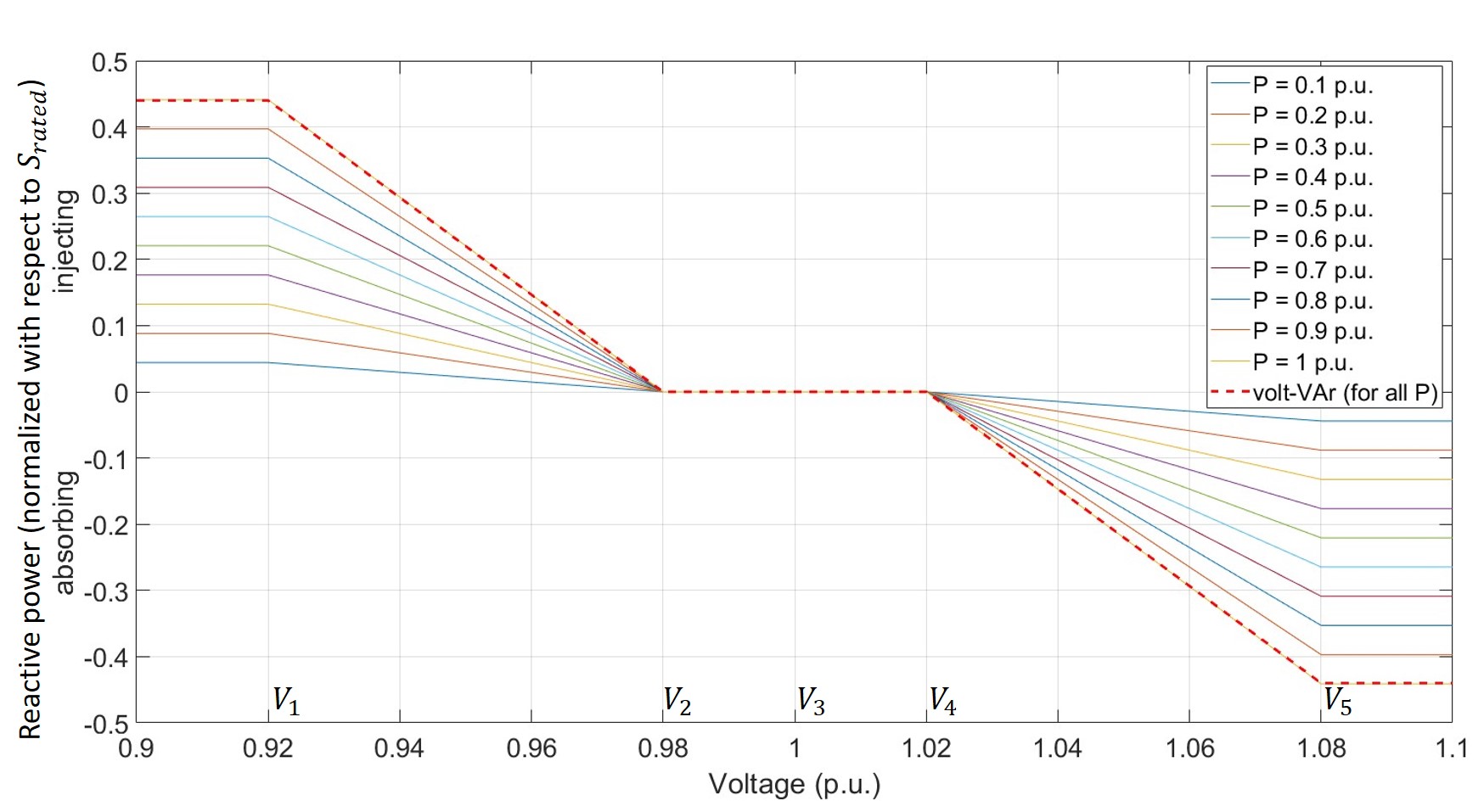}}
\caption{Reactive power vs. voltage relationship for the proposed volt-PF ver. 2 control scheme at different active power levels.}
\label{fig22}
\vspace*{-5mm}
\end{figure}

 The resulting $Q$ vs $V$ curves for volt-PF ver. 2 at various active power conditions are shown in Fig. \ref{fig22} where the linear relation between $Q$ and $V$, and the desired variation in $Q$ as the active power changes are apparent.  It can be seen that for operation at the rated active power, volt-PF and conventional volt-VAr curves are identical, and as the active power reduces the corresponding $Q$ also reduces proportionately in the volt-PF case while it remains unchanged in the volt-VAr case.  It may be noted that the volt-PF ver. 2 can be thought of as volt-VAr mode but with the values +0.44 and -0.44 scaled by the operating active power, which leads to fair $Q$ allocation and a narrow range of PF.

\par The implementation of volt-PF ver.2 is notably straightforward even when compared to the original volt-PF approach, as it can be treated as a derived version of volt-VAr involving just a scaling factor dependent on $P$. 

\section{Conclusion}
\par This paper has introduced a novel distributed voltage regulation approach, namely volt-PF, for DERs in high penetration distribution grids. By integrating power factor as a regulation technique, the method effectively controls the reactive power output of DERs based on not only the PCC voltage but also the inverter's operating active power. Leveraging these two variables, the proposed volt-PF achieves effective feeder voltage regulation while ensuring fair and equitable reactive power support from all the DERs with minimal variation in the inverter power factor. 
\par The performance of the proposed volt-PF control is evaluated and compared with volt-VAr control method under two different settings corresponding to IEEE 1547-2018 and Hawaiian Electric SRD v1.1 standards.  Extensive static and dynamic simulations on a real distribution feeder with high PV penetration are used for this comparison and validation of volt-PF control method. These results demonstrate that the new control mode significantly outperforms the volt-VAr control in both settings by efficiently regulating the grid voltage while reducing reactive power consumption by 17\% and 28.5\%, respectively. Unlike the volt-VAr method, the volt-PF control ensures fair reactive power support from the inverters based on their active power generation, achieved through tightly regulating the inverter power factor, which consistently remains above 0.9 (compared to PF as low as 0.507 for volt-VAr control in Hawaiian Electric SRD v1.1 in the feeder studied). These improvements also contribute to reduced loading of distribution transformers and reduced active power loss in the feeder.
\par Two different versions of volt-PF are described. In the first version, the power factor is a piecewise linear function of the voltage and the reactive power is a non-linear function, while in the second version, the reactive power is a piece-wise linear function of the voltage and proportional to the operating active power.  Both schemes are simple and straightforward to implement with only minor code changes. Hardware changes are not needed since most DER inverters support the constant PF mode, and hence already have the required sensors for implementing any commanded PF.  Similar to volt-VAr mode, the volt-PF is a fully distributed control method without the need for any communication, while additionally offering the advantages of fair reactive power support burden and uniformly high operating PF for all DER inverters.

\section*{Acknowledgments}
The authors would like to thank Poundra, LLC for their great support for the hardware design and the technical documents of the EID prototypes used for the commercial DER control, Arizona Public Service for providing various inputs and for their work towards the project. The material presented in this paper is based upon work supported by the U.S. Department of Energy’s Office of Energy Efficiency and Renewable Energy (EERE) under the Solar Energy Technologies Office Award Number DE-EE0008773.

\section{Biography Section}

\vspace*{-10mm}
\begin{IEEEbiographynophoto}{Madhura Sondharangalla}(Student Member, IEEE) received his B.S. degree in electronics and telecommunication engineering from the University of Moratuwa, Moratuwa, Sri Lanka, in 2017. He is currently pursuing the Ph.D.degree in electrical engineering in Arizona State University, Tempe, AZ, USA. 

\end{IEEEbiographynophoto}
\vspace*{-10mm}
\begin{IEEEbiographynophoto}{Dan Moldovan}(Student Member, IEEE) is a PhD student in electrical engineering at Arizona State University. He received B.S. degree from the University of Portland, Portland, Oregon, USA, in 2020 and the M.S. degree in electrical engineering from Arizona State University, Tempe, AZ, USA, in 2023. His current research interests are renewable energy integration, smart grids, and cybersecurity.

\end{IEEEbiographynophoto}
\vspace*{-10mm}
\begin{IEEEbiographynophoto}{Raja Ayyanar}(Fellow, IEEE) received the M.S. degree in electrical engineering from the Indian Institute of Science, Bengaluru, India, in 1995, and the Ph.D. degree in electrical engineering from the University of Minnesota, Minneapolis, MN, USA, in 2000. He is currently a Professor with Arizona State University, Tempe, AZ, USA. 

\end{IEEEbiographynophoto}

\vfill

\end{document}